# Real time state monitoring and fault diagnosis system for motor based on LabVIEW[1]

S.Q. Liu, Z.S. Ji, Y Wang, Z.C. Zhang

*Abstract*—**Motor is the most widely used production equipment in industrial field. In order to realize the real-time state monitoring and multi-fault pre-diagnosis of three-phase motor, this paper presents a design of three-phase motor state monitoring and fault diagnosis system based on LabVIEW.**

**The multi-dimensional vibration acceleration, rotational speed, temperature, current and voltage signals of the motor are collected with NI cDAQ acquisition equipment in real time and high speed. At the same time, the model of motor health state and fault state is established. The order analysis algorithm is used to analyze the data at an advanced level, and the diagnosis and classification of different fault types are realized. The system is equipped with multi-channel acquisition, display, analysis and storage. Combined with the current cloud transmission technology, we will back up the data to the cloud to be used by other terminals.**

## I. Introduction

In today's manufacturing industries, mechanical and electromechanical systems are driven by electric motors on the premises [1]. Any small fault occurred in a motor will led to complete motor failure if not addressed in time[2].In recent years, with the development of sensor technology and signal processing technology, such as filtering technology, spectrum analysis technology and so on, the motor fault diagnosis is becoming more and more perfect.

In order to ensure real-time data access and data integrity, the system is built with two layers producer and consumer frame.

The key requirements of the state monitoring and fault diagnosis system are listed as follows:

1. Measure the real-time parameters of the motor, including triaxial acceleration, rotational speed, temperature, working current and voltage, at 1KSPS to 25 KSPS sampling rate per channel.
2. Process the raw data. After filtering and amplifying, the data can be transformed into actual physical quantities.
3. Display the actual parameters and state of the motor on the monitoring interface of the Industrial Personal Computer (IPC).
4. Analyze the critical state of each parameter and the specific fault source of motor by multiple analysis algorithm.
5. Archive all of data in the MySQL database on local and Ali cloud servers at intervals of 10ms.

## II. System Structure

According to the above requirements, the state monitoring and fault diagnosis system includes the following functions:

- Real-time data acquisition
- Data processing and analysis
- State monitoring
- Data archiving and managing

The system's architecture is as shown in Fig. 1.The system acquires data of sensors on healthy motor and faulty motor through NI cDAQ acquisition device. After processing and analysis, the data will be displayed on the monitoring interface of the IPC. All parameters and state information are stored in the MySQL database on local and Alibaba cloud server.

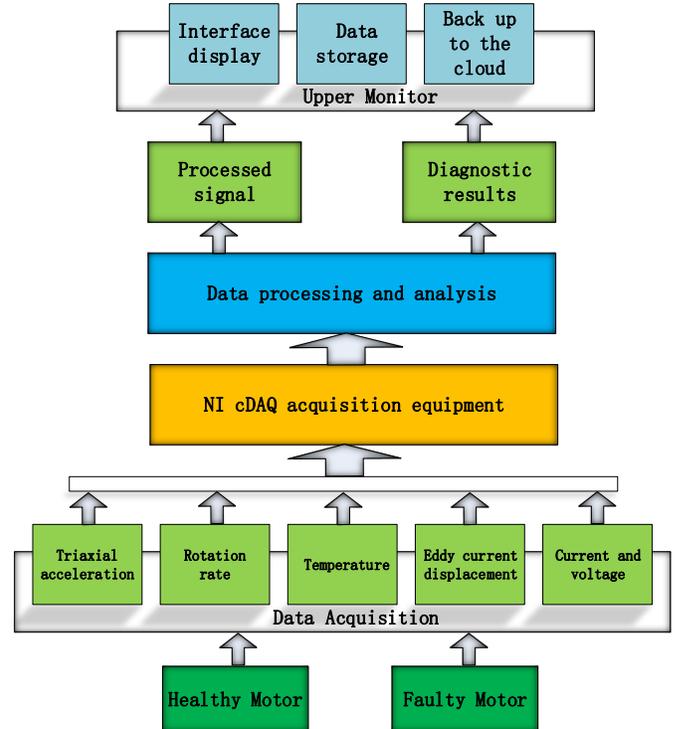

Fig. 1. System architecture

### A. Real-time data acquisition

The system acquires the signals of the underlying sensors from NI cDAQ acquisition equipment in real time, and transmits the signals to the IPC based on the TCP/IP protocol

This work is supported by the National Natural Science Foundation of China under Grant No.11505239.

S.Q. Liu is with the Institute of Plasma Physics, Chinese Academy of Sciences, Hefei, Anhui, China; and also with University of Science and Technology of China, Hefei, China (telephone: 86-551-65592397; e-mail: shqliu@ipp.ac.cn).

Z.S. Ji, Y. Wang, Z.C. Zhang are with the Institute of Plasma Physics, Chinese Academy of Sciences, 350 Shushanhu Road, Hefei, Anhui 230031, P. R. China.

[3]. The system samples the vibration signal at a maximum sampling rate of 25KSPS. For other signals, the system samples from 1SPS to 10KSPS.

### B. Data processing and analysis

The system processes the raw data by filtering and amplifying. It uses order analysis to analyze the acceleration and speed signal. Threshold analysis is used to analyze other signals. The final results of the analysis of the motor will be obtained by comprehensive analysis.

### C. State monitoring

The system displays the actual physical quantities and state on the monitoring interface of the IPC. For abnormal parameters, the system will give corresponding prompts through the alarm indicator. At the same time, the user can choose to view the real time parameter waveform.

### D. Data archiving and managing

The system archives all of data in MySQL database at intervals of 10ms. In order to achieve multi-point backup, the system archives data on local and Alibaba cloud server. At the same time, in order to facilitate the user to view the data, the corresponding data management functions are designed.

## III. SYSTEM IMPLEMENTATION

### A. Hardware Structure

Fig. 2 shows the main hardware structure of the state monitoring and fault diagnosis system.

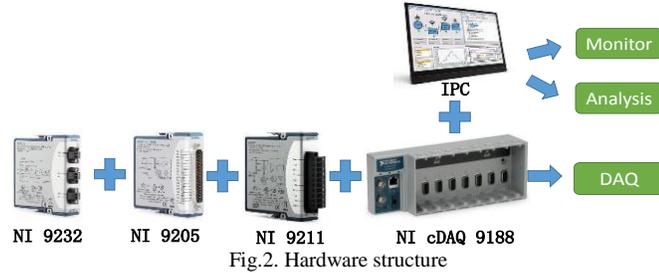

Fig.2. Hardware structure

1. NI cDAQ 9188 is 8-slot Ethernet chassis, which is used to communicate with the IPC [4].
2. NI 9232 is a 3-channel, 102.4 KSPS, ±30V, C series sound and vibration input board [5].It is used to obtain triaxial acceleration signals.
3. NI 9205 is a 16-bit, 32-channel C series voltage input card [6] for speed and current voltage signals.
4. NI 9211 is a 4-channel, ±80 MV C series temperature input board [7] for obtaining thermocouple temperature signals.
5. IPC is an industrial control computer for data processing and status monitoring.

### B. Software Structure

The software is mainly based on LabVIEW, and its environments are listed as follows:
1. LabVIEW 2015
2. MySQL 5.7.20
3. Alibaba cloud server (Centos 7.3)

As shown in fig.4, the software system is divided into three modules: data acquisition module, data processing module and state monitoring and storage module. Data flow is based on a two-tier producer and consumer framework, which ensures both real-time data collection and data integrity.

Data-acquisition module is used to acquire data from the NI cDAQ acquisition equipment and sends raw data to data-process module. It analysis DAQ info, configures acquisition setting and acquires data [8].The sampling rate can be adjusted from 1KSPS to 25KSPS according to the acquisition configuration information. In order to realize data integrity, there is a buffer area between the data-acquisition module and the data-process module.

Data-process module is used to preprocess and analyze data. It uses Kalman filter to process the raw data [9]. In order to obtain the real physical value of the raw data, the data will be amplified in a certain proportion. Data analysis consists of two parts, primary analysis and advanced analysis. The primary analysis is mainly the threshold analysis, which judges the state of the motor by the upper limit and lower limit of the parameters. Advanced analysis is mainly order analysis algorithm.

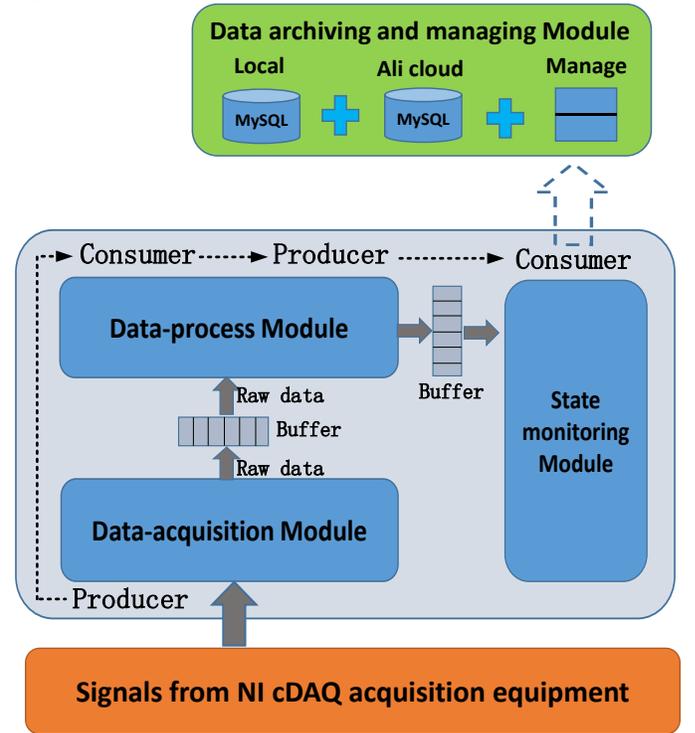

Fig.3. Software structure

The order spectrum is used to locate the fault source of the motor. First of all, the crankshaft is sampled at equal angles. In order to obtain the resampling time point [10], it is assumed here that the reference axis rotation angle θ can be represented in the following form:

$$\theta(t) = b_0 + b_1 t + b_2 t^2 \qquad (1)$$

Where: $b_0$, $b_1$, $b_2$ is the undetermined coefficient; t is the time point.

At the same time interval sampling signal, three pulse time points t and rotation angle increment O which have been achieved in turn are brought into (1).It gets

$$\begin{pmatrix} O \\ \Delta O \\ 2\Delta O \end{pmatrix} = \begin{pmatrix} 1 & t_1 & t_1^2 \\ 1 & t_2 & t_2^2 \\ 1 & t_3 & t_3^2 \end{pmatrix} \begin{pmatrix} b_0 \\ b_1 \\ b_2 \end{pmatrix} \quad (2)$$

After finding b, we take it into (1) and get

$$t = \frac{1}{2b_2}[4b_2(\theta - b_0) + b_1^2 - b_1] \quad (3)$$

According to the above formula, the time points corresponding to the constant angle increment when sampling at equal angles can be obtained. After the acceleration signal is interpolated and Fast Fourier Transform (FFT) according to the time point, the order spectrum will be obtained. The system analyzes the difference of the amplitude of each order when the rotational speed changes, and gets the fault information of the motor finally. Fig.4 is a flow chart of order analysis algorithm.

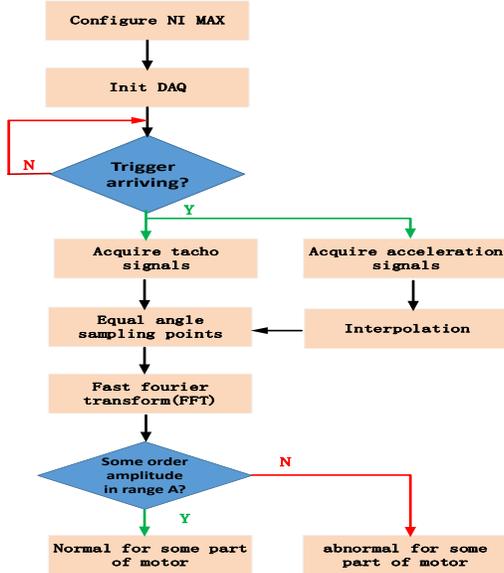
Fig.4. Work flow of order analysis algorithm

State monitoring module is in front of the system. Based on the two-tier producer and consumer model, the actual physical quantities and state of motor is displayed to the front end of the monitor interface in real time by data flow. For abnormal parameters, the system will give corresponding prompts through the alarm indicator. Meanwhile, the system will also give specific alarm information on the interface. The user can choose to view the real time parameter waveform of data.

Data archiving and managing module mainly consists of two parts. In order to achieve multi-point backup, the system archives data in MySQL database on local and Alibaba cloud server. The system also adds the function of managing the database. Users can view the system's historical data in the form of a chart. The system remotely connects to the MySQL database on the cloud server through the LabSQL toolkit. At present, the database is mainly divided into four tables.Fig.5 is the system remote database table structure.

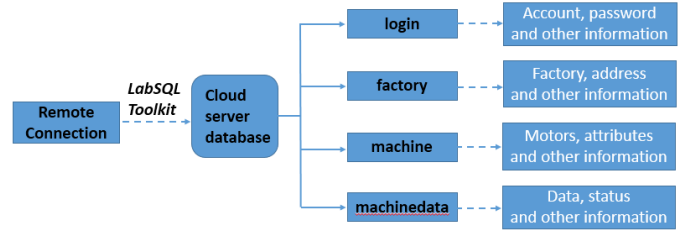
Fig.5. Remote database table structure

## IV. RESULTS

### A. Real-time data monitoring results

The state monitoring and fault diagnosis system was tested in March 2018. The system has been running steadily for two months in the pilot phase. The test results show that it meets the requirements of real-time data acquisition and monitoring, and has the characteristics of persistence and robustness at the same time.

Fig.6 shows the monitoring interface of the system. It includes sensor data, state, and alert information.

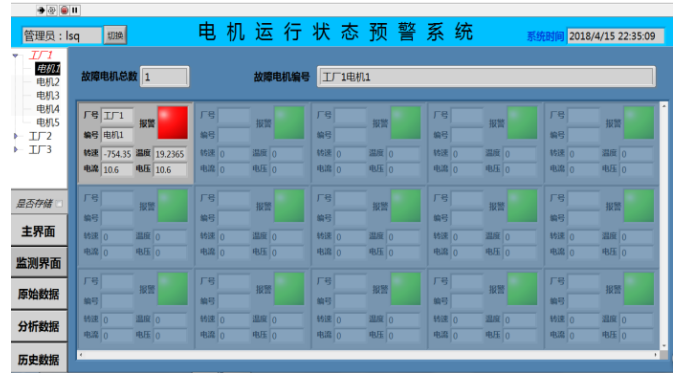
Fig.6. The monitoring interface of the system

### B. Fault diagnosis results

Fig.7 shows the order spectrum of a healthy motor and a faulty motor. From the figure, we can see that the amplitude of the faulty motor at the 10th and 14th orders is much higher than that of the healthy motor. This reflects a fault feature of the motor from the side. It provides a good reference for users.

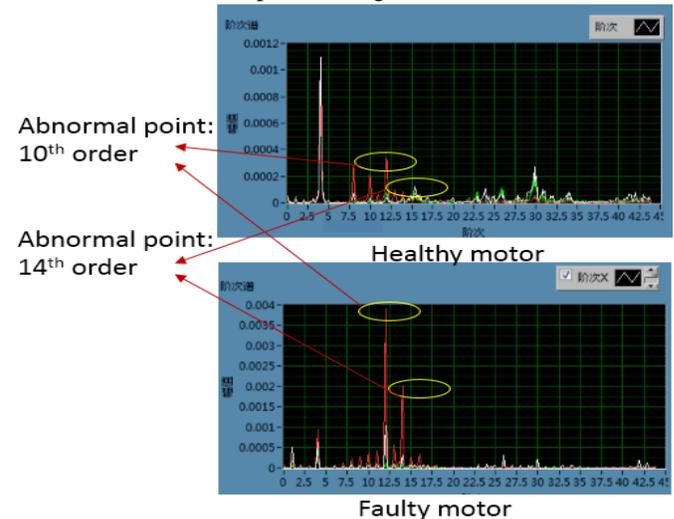
Fig.7. Order spectrum contrast graph

### C. Data managing results

Fig.8 shows the interface of the data management. In order to meet the needs of users to view historical data, multiple search functions are designed. As shown in the figure, users can retrieve data by setting query time and query parameters. The system also provides export data to Excel table function. Data management features provide users with a great deal of convenience.

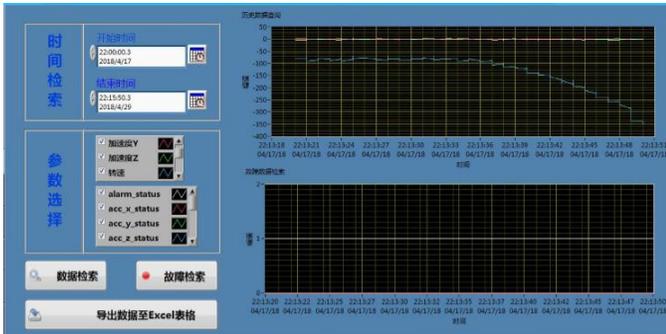

Fig.8. the interface of the data management

## V. Summary

In this paper, we have presented a system based on LabVIEW. This system realizes multi-functions such as data acquisition, data processing and analysis, and state monitoring and archiving. In order to achieve multi-point backup, the system uses Alibaba cloud server for data cloud storage. After two months of testing, the system basically meets the requirements of the system design. The function of real-time data acquisition and data processing and analysis runs stably.

Next work, we will continue to perfect system functions and add multiple analysis algorithms.

## Acknowledgment

This work is supported by the National Natural Science Foundation of China under Grant No.11505239.

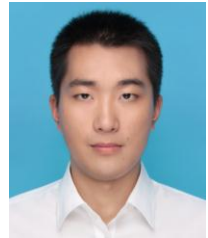

**S.Q. Liu** received the B.S. degree in measurement and control technology and instruments from Anhui University of Technology, Maanshan, China, in 2014 and the M.S. degree in computer technology from University of Science and Technology of China, Hefei, China, in 2018. He is currently pursuing the Ph.D. degree in computer application technology at University of Science and Technology of China, Hefei, China.

From 2015 to 2018, He was a student with Institute of Plasma Physics, Chinese Academy of Sciences, Hefei, China. His research interest includes data acquisition and automatic control in the industrial field.

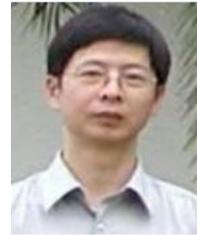

**Z.S. Ji** was born in 1963. He received the B.S. degree from the United University of Hefei , China, in 1985.

Since 1985, he has been engaged in the control and data acquisition systems of the Institute of Plasma Physics of the Chinese Academy of Sciences. As the subject leader, he has participated in several 973 projects of the Ministry of Science and Technology and the National Natural Science Foundation of China. As a visiting scholar, he worked at Kyushu University and NIFS Institute in Japan.

Now he is responsible for the total control system and safety chain system, hardware and signal conditioning of the National Science and Engineering EAST device. Over the years, he has published more than 40 papers on control and obtained more than 10 national invention patents.